\documentclass[twocolumn]{aastex631}

\usepackage{graphicx}
\usepackage[FIGTOPCAP]{subfigure}
\usepackage{amsmath,booktabs,threeparttable,url, bm}
\usepackage{CJKutf8}
\newcommand{\cntext}[1]{\begin{CJK*}{UTF8}{bsmi}#1\end{CJK*}}

\usepackage{natbib}
\received{\today}
\shorttitle{M-dwarf companion}
\shortauthors{Pan et al.}

\graphicspath{{./}{figures/}}

\begin{document}

\title{Supernova Ia Remnants with M dwarf surviving companions}

% Author affiliation commands
\newcommand*{\NTHUP}{Department of Physics, National Tsing Hua University, Hsinchu 30013, Taiwan}
\newcommand*{\NTHUA}{Institute of Astronomy, National Tsing Hua University, Hsinchu 30013, Taiwan}
\newcommand*{\CICA}{Center for Informatics and Computation in Astronomy, National Tsing Hua University, Hsinchu 30013, Taiwan}
\newcommand*{\CTC}{Center for Theory, Computation, and Data-Science Research, National Tsing Hua University, Hsinchu 30013, Taiwan}
\newcommand*{\IFF}{Instituto de F\'{i}sica Fundamental, Madrid, Spain}
\newcommand*{\ICC}{Institut de Ci\`{e}ncies del Cosmos, Barcelona, Spain}
\newcommand*{\IAC}{Instituto de Astrof\'{i}sica de Canarias, Tenerife, Spain}
\newcommand*{\ULL}{Universidad de La Laguna, Dept. Astrof\'{i}sica, Tenerife, Spain}

% %% Note that the corresponding author command and emails has to come
% %% before everything else. Also place all the emails in the \email
% %% command instead of using multiple \email calls.
% \correspondingauthor{August Muench}
% \email{greg.schwarz@aas.org, august.muench@aas.org}

\author[0000-0002-1473-9880]{Kuo-Chuan Pan (\cntext{潘 國全})}
\email{kuochuan.pan@gapp.nthu.edu.tw}
\affiliation{\NTHUP} \affiliation{\NTHUA}

\author[0000-0001-9046-4420]{Pilar Ruiz-Lapuente}
\affiliation{\IFF} \affiliation{\ICC}

\author[0000-0002-0264-7356]{Jonay I. Gonz\'{a}lez Hern\'{a}ndez}
\affiliation{\IAC} \affiliation{\ULL}

\begin{abstract}
  
We study the possibility that Type Ia supernovae might be produced
by binary systems where the companion of the exploding white dwarf is
an M-dwarf star. Such companion would appear as a runaway star, retaining its
pre-explosion orbital velocity along with a kick imparted by the supernova
ejecta. It might be rapidly rotating, from being tidally locked with
the white dwarf prior to explosion in a very close binary. For this study, we
perform
a series of multidimensional hydrodynamic simulations to investigate the
interaction between M-dwarf companions and SN ejecta, followed by
post-impact stellar evolution modeling using the MESA code.
Our initial models in the 3D simulations had high spin angular momenta
and the effects of magnetic braking have been included. They very
significantly reduce the final rotation. 

A surviving companion candidate, MV-G272, has recently been discovered in the
supernova remnant G272.2-3.2, which is an 8.9$\sigma$ proper motion
outlier, although being slowly rotating.
Our results show that the properties of this companion (luminosity, effective
temperature, surface gravity) can be reproduced by
our post-impact M-dwarf models. The slow rotation, which is a common
characteristic with several proposed hypervelocity SN companions, can be
explained by 
magnetic braking during the post-impact evolution, thus supporting
the possibility that the MV-G272 star is the  surviving companion of
the Type Ia supernova now found as G272.2-3.2 SNR. 

\end{abstract}


\section{Introduction}

Type Ia supernovae (SNe Ia) are powerful cosmological probes, as proved by the discovery of the accelerated expansion of the universe \citep{1998AJ....116.1009R, 1997ApJ...483..565P} and by their current role in the exploration of the evolution of dark energy \citep{2023arXiv231112098R, 2025arXiv250306712D}. 
They are thermonuclear explosions of carbon-oxygen white dwarfs (WDs), members of close binary systems. Mass accretion from their companion stars induces the explosions. There is no evidence, at present, on the nature of that stellar companion, so they may as well be a main-sequence, subgiant, red giant, AGB star, a He star, or another WD. The mass-accretion mechanism itself is determined by the nature of the companion: Roche-lobe overflow, stellar wind, merging, or collision with another WD. Depending on that, thermonuclear burning may be started and propagated in different ways (central deflagration, delayed detonation, surface detonation followed by central detonation).

The evolutionary scenario first proposed to produce a Type Ia supernova from a WD \citep{1973ApJ...186.1007W, 1982ApJ...253..798N}, called the single-degenerate scenario (SD), assumes the growth in mass of a WD up to the Chandrasekhar mass by accretion of material from a non-degenerate companion through Roche-lobe overflow. Rapid contraction then explosively ignites the central layers of the WD (made of C+O) when reaching a critical density. The explosive thermonuclear burning would start as a deflagration, or a deflagration that evolves into a detonation farther from the center (delayed detonation). The companion, although hit by the SN ejecta, can survive \citep{2000ApJS..128..615M, 2010ApJ...715...78P, 2012ApJ...750..151P, 2012A&A...548A...2L, 2022ApJ...933...38R, 2023RAA....23h2001L}.

In the SD scenario, the companion star of the exploding WD is ejected with the orbital velocity plus the kick due to the impact of the SN material. 
Velocities in the hundreds of km s$^{-1}$ should be expected in this case \citep{2012ApJ...750..151P}. 
Post-explosion evolution of different possible SN Ia companions has been calculated by \cite{1975ApJ...200..145W, 2012ApJ...750..151P, 2014ApJ...792...71P, 2019ApJ...887...68B, 2021A&A...654A.103L, 2022ApJ...933...38R, 2024ApJ...973...65W, 2025ApJ...989...72W}.

On the contrary, if the companion star were another WD \citep{1984ApJ...277..355W, 1984ApJS...54..335I}, which is called the double-degenerate (DD) scenario, the companion should be destroyed by the mass-transfer process itself. In this model, the orbit of the two WDs decreases due to the emission of gravitational waves. Eventually, tidal interaction disrupts the companion (the less massive WD or secondary) and its debris merges with the primary WD, increasing its mass towards the Chandrasekhar limit. 
Recent modelling, however, has shown that the explosion can be produced differently in the ``violent merger'' DD scenario, the interaction of the debris of the secondary WD after its collision with the primary can trigger a prompt detonation while the merger is still ongoing, given as a result a SN explosion whose characteristics have been seen in some observed SN Ia \citep{2010Natur.463...61P, 2012ApJ...747L..10P, 2015ApJ...807..105S}.

In the core-degenerate (CD) model, a WD merges with the electron-degenerate core of an AGB star \citep{2013IAUS..281...72S, 2014MNRAS.437L..66S, 2019NewAR..8701535S}. In this case, the WD is engulfed by the envelope of the companion, and friction leads to merging. 
If there is sufficient delay between merging and explosion (the merger to explosion delay, MED, model; \citealt{2019NewAR..8701535S, 2022RAA....22c5025S}), the exploding object may have reached the Chandrasekhar mass and undergo central ignition, developing into a delayed detonation. 
%No companion is left in this scenario. 

There is a new turn on the outcome of the DD scenario: the so-called Dynamically-Driven Double-Degenerate Double-Detonation model (D$^{6}$; see \citealt{2025ApJ...982....6S}, for instance, for a recent account), where the explosion happens when the less massive WD has shed only a part of its mass, the rest being still gravitationally bound. 
If the secondary were a He WD, a He shell detonation could be triggered by the violent impact of the accretion stream on the primary WD.
%by the He accumulated, by steady mass transfer, on top of the primary (made of C+O). 
The ensuing compression can induce a second detonation close to the center. Due to the extreme closeness of the two WDs at the time of the SN Ia explosion, in this model, the surviving WD should be ejected at very high velocities. This can be the origin of hypervelocity stars (WDs with velocities $ > 1000$ km
s$^{-1}$) with peculiar characteristics. It has also been proposed that high-velocity stars could result from the only partial burning of a WD in a SN Iax explosion \citep{2019MNRAS.489.1489R}.

A few hypervelocity stars have been found that clearly cannot have been accelerated at the Galactic center \citep{2018ApJ...865...15S, 2023OJAp....6E..28E}, which points to the D$^{6}$ scenario for their production. Only for a couple of the suggested discovered companions from the D6, an idea of the rotational velocity is known. It is a small velocity when compared with the translational velocity of those stars. Thus, they are not fast rotators. 
On the other hand, in the SD scenario, if the two stars are tidally locked at the time of the explosion, the surviving companion should be rotating fast. There are, however, mechanisms slowing down the pre-explosion rotation, mostly related to mass loss from the companion, during the explosion and later on \citep{2013A&A...554A.109L, 2014ApJ...792...71P}.

Among the hypervelocity stars thought to originate from SN Ia explosions, US 708 is a particularly compelling example. As a rapidly rotating helium-rich subdwarf O star (sdO), it likely survived a single-degenerate SN Ia explosion \citep{2015Sci...347.1126G, 2017A&A...601A..58Z}.
Its extremely high velocity ($> 1000$ km/s) and current location unbound from the Milky Way suggest a violent ejection mechanism consistent with thermonuclear detonation in a close binary. US 708 thus exemplifies how some post-SN Ia survivors can be fast rotators and hypervelocity stars.

A possible surviving companion of a SN Ia has recently been discovered in the Galactic supernova
remnant (SNR) G272.2-3.2 \citep{2023ApJ...947...90R}, using the data from the Gaia satellite. It has been named MV-G272. The star is a 8.9$\sigma$ outlier in proper motion, and its trajectory places it at the center of the remnant at the time of the SN explosion, $\sim$ 7500 yr ago.
MV-G272 is a M1-M2 dwarf with a large space velocity and is slowly rotating. Here, we investigate the way a M dwarf star may show the characteristics of MV-G272 (space and rotational velocities, mass, luminosity, effective temperature, and surface gravity) after being hit by the ejecta of a SN Ia and has been cooling down for the SNR age afterwards.

In the standard single-degenerate (SD) scenario for SNe Ia, M dwarfs are generally not considered likely binary companions due to their long delay time before initiating stable Roche-lobe overflow. However, \citet{2012ApJ...758..123W} proposed a novel SD channel in which WD+M-dwarf binaries could produce SNe Ia if they emerge from a common-envelope phase and mass transfer is aided by the “magnetic bottle” effect of a fully convective, low-mass M-dwarf companion.
In this paper, we investigate in depth the outcome of SNe Ia explosions with low-mass main-sequence companions. We explore the evolution with time after the explosion in the H-R diagram. We determine if MV-G272 can be the companion of the SNIa that gave rise to the remnant G272.2-3.2.

\section{Observations}

The Gaia EDR3 was used to obtain the proper motions, parallaxes, and photometry of stars within a circle of 11' radius on the sky around the centroid of the SNR G272.2-3.2 \citep{2023ApJ...947...90R}. The parallaxes were selected according to the distance to the SNR. That made a sample of 3082 stars.

From the proper motions, star MV-G272 appeared as an extreme outlier, at 5.8$\sigma$ from the mean in R.A. and 8.4$\sigma$ in DEC. The total proper motion is at 8.9$\sigma$ above the mean. From its parallax, the star has a tangential velocity of $v_{\rm tan} = 239^{+181}_{-70}$ km s$^{-1}$, which is 5.4$\sigma$
above the mean for the sampled stars. Gaia does not provide radial velocities, but those were obtained from the spectra used to determine the spectral class and stellar parameters of MV-G272. The radial velocity was $v_{\rm R} = 92.5\pm0.5$ km s$^{-1}$, which gave a total velocity of $v_{\rm tot} = 256^{+181}_{-70}$ km s$^{-1}$.

When extrapolating the measured proper motion backwards, until the time of the SN explosion, the position of MV-G272 becomes very close to the centroid of the SNR (see Fig. 5 in \citealt{2023ApJ...947...90R}). That further points to this star as a likely companion of the SN.

Spectra were obtained with the MIKE spectrograph on the 6.5 m Clay telescope and with the Goodman spectrograph on the 4.1 m SOAR telescope.
Apart from measuring the radial velocity, they were used to determine the spectral type, luminosity class, stellar parameters, surface abundances, and rotational velocity of MV-G272. The obtained characteristics are shown in Table 1. The rotational velocity found is $v_{\rm rot} \sin i \leq 4$ km s$^{-1}$, in contrast with the total velocity. This point is discussed in Section 5.2.

\begin{table}
  \centering
  \caption{Characteristics of the star MV-G272}
    \begin{tabular}{lccc}
      \hline
      \hline
      Spectral type & & & M1-M2 \\
      Luminosity class & & & V \\
      $T_{\rm eff}$ (K) & & & 3600--3850 \\
      log $g$ & & & 4.46$^{+0.10}_{-0.11}$ \\
      $[Fe/H]$ & & & -0.32$\pm$0.04 \\
      log($L(L_{\odot}$) & & & -1.54/-1.39 \\
      M ($M_{\odot}$) & & & 0.44-0.50 \\
      R ($R_{\odot}$) & & & 0.446-0.482 \\
\hline
\end{tabular}
\end{table}

As mentioned in the Introduction, the hypervelocity or high-velocity stars proposed as surviving companion candidates of SNe Ia do not show fast rotations either. Three stars have their rotational velocities measured (LP 398-9, LP 40-365, and J0546+0836; see Table 2). 
In LP 398-9, a periodic 15.4 h signal in the UV and optical light curves has been detected by \cite{2022MNRAS.512.6122C} and interpreted as surface rotation. LP 40-365 has been attributed to a partially burned WD in a Type Iax SN.
A photometric variation corresponding to a 5.8\% in the UV flux, with a period of 8.914 h has been measured by \cite{2021ApJ...914L...3H}. The rotational velocity of J0546+0836 has been estimated by \cite{2023OJAp....6E..28E}, from the broadening of the $\lambda\lambda$5801, 5012 $\AA$ doublet of C IV. 
In Table 2, their periods, radii, linear and rotational velocities, together with those of MV-G272, as well as the proportion of the rotational to linear velocities, expressed in percentages, are shown.  

\begin{deluxetable*}{lcccccc}
\tablewidth{0pt} 
\label{tab:observations}
\tablecaption{Linear and rotational velocities of proposed surviving companions of SNe Ia}

\tablehead{
Object & $P$ & $R$ & $v$ & $v_{\rm rot}$ sin $i$ & \%  & Ref. \\
& [hr] & [$R_{\odot}$] & [km s$^{-1}$] & [km s$^{-1}$] &  &  \\
}
\startdata
MV-G272 & 66.92 & 0.564 & 363 &  $\leq$ 4 & $\leq$ 1.1 & \cite{2023ApJ...947...90R} \\
 D6-1 &         &        & 2045$^{+251}_{-187}$ & $\leq$ 20 & $\leq$ 0.97 & Shen (2025), private communication \\
 LP 398-9 (D6-2) & 15.4 & 0.20$\pm$0.01 & 1151$^{+59}_{-51}$ & $\leq$ 15.78 &
 $\leq$ 1.37 & \cite{2022MNRAS.512.6122C} \\ 
LP 40-365 (GD 492) & 8.914 & 0.16$\pm$0.01 & 837$^{+5}_{-5}$ & 21.80 & 2.6 & \cite{2021ApJ...914L...3H} \\
J0546+0836 &  0.33$^{*}$  & 0.051$^{+0.029}_{-0.021}$ & 1699$^{+670}_{-390}$ & 180 & 10.6 & \cite{2023OJAp....6E..28E}
\enddata
\tablenotetext{}{{\bf Notes}
  $^{*}$ Calculated from
  $P_{\rm rot} \simeq$ 20 min $\times \left({R\over0.05 R_{\odot}}\right)$ sin $i$}
\end{deluxetable*}

%In the $D^6$ scenario \citep{2018ApJ...865...15S, 2025ApJ...982....6S},}
There is no evidence of rotational broadening in the three hypervelocity white dwarfs found by \cite{2018ApJ...865...15S} (D6-1, D6-2 and D6-3).
For D6-1 the upper limit is consistent 
with $<$ 20 km s$^{-1}$ (Shen 2025, private communication).
An upper limit of 20 km s$^{-1}$ has been set. That means less than 1\% of the linear velocity.
In LP 398-9 (D6-2), a periodic 15.4~h signal in the UV and optical light curves has been detected by \cite{2022MNRAS.512.6122C} and interpreted as surface rotation. Although that is not clear (it might be due to some sort of circumstellar gas/disk), they attribute a radius of 0.20$\pm$0.01 $R_{\odot}$ (see their Table 1). 
For such radius, a rotation period of 15.4 h means a $v_{\rm rot}$ sin $i$ = 15.776 km s$^{-1}$ only, that is a 1.37 \% of its linear velocity.

LP 40-365 has been attributed to a partially burned WD in a Type Iax SN.
A photometric variation corresponding to a 5.8\% in the UV flux, with a period of 8.914 h has been measured by \cite{2021ApJ...914L...3H}. This was based on the light curves obtained by the Transiting Exoplanet Survey Satellite (TESS), the Hubble Space Telescope, and the Wide-field Infrared Survey Explorer (WISE) spacecraft. 
\cite{2019MNRAS.489.1489R} had determined the parameters of the star, finding a radius $R = 0.16\pm0.01 R_{\odot}$. From that, a rotational velocity $v_{\rm rot}$ sin $i$ = 21.80 km s$^{-1}$, that is a  2.6 \% of its linear velocity.

The rotational velocity of J0546+0836 has been estimated by \cite{2023OJAp....6E..28E}, from the broadening of the $\lambda\lambda$5801, 5012 \AA \ doublet of C IV. Comparison with star HE 1429-1209 shows the doublet in J0546+0836 to be both stronger (with an equivalent width of 7 \AA, compared to $\sim$ 3 \AA \ for HE 1429-1209) and broader. If the observed broadening is due to rotation, this would imply $v_{\rm rot}$ sin $i \simeq$ 180 km s$^{-1}$, which implies a rotation period $P_{rot} = \frac{2\pi R \sin i}{v_{\rm rot} \sin i} \simeq$ 20 min $\times \left({R\over0.05 R_{\odot}}\right)$ sin $i$.
It is not excluded, however, that processes besides rotation dominate the emission-line broadening. Here, the rotational velocity $v_{\rm rot}$ is a 10.6 \% of the linear velocity.

% =================================================================
%
%   Simulation methods
%
%  ================================================================

\section{Numerical methods} \label{sec:methods}
% Method
The numerical methods and setup are essentially similar to those implemented in \cite{2010ApJ...715...78P, 2012ApJ...750..151P, 2012ApJ...760...21P, 2013ApJ...773...49P, 2014ApJ...792...71P, 2022ApJ...933...38R, 2023ApJ...949..121C}, but with improvements and adjustments that will be described later. We first use the stellar evolution code {\tt MESA} to construct low-mass companion models, followed by the hydrodynamics code {\tt FLASH} to perform 2D and 3D simulations of the supernova's impact on the companion star. Finally, we incorporate information on mass stripping, supernova heating, and angular momentum loss from these hydrodynamics simulations into {\tt MESA} to conduct the subsequent post-impact evolution of the surviving companions. 

%
% Figure 1
%

\begin{figure}
\epsscale{1.2}
\plotone{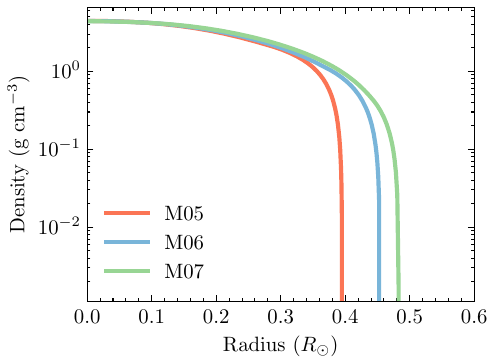}
\caption{Density profiles of our considered companion models. Different colors represent models with different initial masses.}
\label{fig:models}
\end{figure}

\subsection{Companion Models}

We use the stellar evolution code {\tt MESA}, version r10398 \citep{2011ApJS..192....3P,2013ApJS..208....4P, 2015ApJS..220...15P, 2018ApJS..234...34P, 2019ApJS..243...10P, 2023ApJS..265...15J}, to construct three low-mass companion models with masses $M_c= 0.5, 0.6$, and $0.7 M_\odot$ and with metallicity $z=0.02$. 
Since the detailed binary evolution in the single-degenerate scenario prior to a SN~Ia explosion remains elusive, it is possible that the binary system underwent a common-envelope stage, leading to a low-mass binary companion prior to the explosion \citep{2012ApJ...758..123W}.
We assume that these companion models have the same structure as at their Zero-Age Main Sequence (ZAMS) stage because their MS lifetimes are much longer than the delayed time of a SN~Ia, and we neglect binary interactions during the binary evolution phase for simplicity. The corresponding stellar radii of these companion models are $R_c =0.396$, $0.454$, and $0.483 R_\odot$. 
Figure~\ref{fig:models} shows the density profiles of these three companion models.
%$3.10 \times 10^{10}$~cm, $3.92 \times 10^{10}$~cm, and $4.54 \times 10^{10}$~cm.   

\subsection{Supernova Impact}

To perform multi-dimensional hydrodynamics of the impact of SN~Ia ejecta on its companion, we use the publicly available code {\tt FLASH} \citep{2000ApJS..131..273F, 2008PhST..132a4046D}.
We employ the unsplit hydro solver with {\tt PARAMESH4\_dev} for the adaptive mesh refinement (AMR) to solve the Euler equations. The equation of state used is the Helmholtz equation of state \citep{1999ApJS..125..277T, 2000ApJS..126..501T}. While we ignore the contribution of magnetic fields and nuclear burning, self-gravity is included via the new multiple Poisson solver \citep{2013ApJ...778..181C}, with a maximum angular moment for spherical harmonics $l_{\rm max}=80$.

The density, temperature, and composition of the companion model are mapped onto either axisymmetric 2D cylindrical grids or 3D Cartesian grids in {\tt FLASH}. The compositions considered in this study are hydrogen ($^1$H), helium ($^4$He), and carbon ($^{12}$C). For simplicity, all elements in the companion star heavier than carbon are approximated as carbon. 
In the three-dimensional simulations, the simulation box has a width of 30 times the radius of the companion ($R_c$) in all directions, with the companion located at the center of the box. In two-dimensional simulations, the simulation box dimensions are set to $15 R_c$ in the radial ($r$) direction and $30 R_c$ in the axial ($z$) direction. The companion is positioned on the $z$-axis, $10 R_c$ away from the $+z$ boundary. 

In two-dimensional simulations, we use 8 levels of refinement based on the magnitudes of the second derivatives of density and pressure. Each AMR block contains $8\times8$ cells with $2\times 4$ basic blocks, corresponding to an effective uniform resolution of $2048 \times 4096$. We apply an ``axisymmetric'' boundary condition at the inner $r$-boundary, while an ``outflow'' boundary condition is used for all other boundaries. 
In three-dimensional simulations, we use 7 levels of refinement with the same refinement criteria. Each AMR block contains $8^3$ cells with $4\times 4$ basic blocks. This corresponds to an effective uniform resolution of $2048^3$. To save computing time in 3D calculations, we reduce the refinement levels based on the logarithm distance from the companion's center, but ensure that the region within $r < 0.9 R_\odot$ remains at the highest refinement level.   
The ``outflow'' boundary condition is used in all boundaries in three-dimensional simulations.

Once the companion model is mapped onto the grids, we use the relaxation method described in \cite{2012ApJ...750..151P} and \cite{2022ApJ...933...38R} to dampen the gas velocity for ten dynamical timescales with a damping factor $f_{\rm damp}=0.97$. 
After relaxation, we place the exploding WD at a given binary separation ($A$) with a W7-like explosion \citep{2012ApJ...750..151P}. In this W7 model \citep{1984ApJ...286..644N}, the WD mass is set to $M_{\rm WD}=1.378 M_\odot$, and the explosion energy is $E_{\rm SN}=1.233 \times 10^{51}$~erg, with an average ejecta speed of $v_{\rm SN}=8.527 \times 10^3$~km~s$^{-1}$.

In three-dimensional simulations, we additionally include symmetric-breaking effects, such as the spin and orbital motions of the companion star. The orbital velocity is calculated using Kepler's law, assuming zero eccentricity. A spin-to-orbital ratio of $0.95$ is assumed in this study. We also ensure that the region between the center of the SN explosion and the companion star remains at the highest refinement level.  
The composition of the SN ejecta is approximated as pure nickel ($^{56}$Ni) and is used as a tracer in simulations \citep{2012ApJ...750..151P}.

Table~\ref{tab:models} summarizes all the 2D and 3D models we considered in the hydrodynamics simulations. 
Note that in the standard single-degenerate scenario, the binary companions are expected to experience Roche-lobe overflow at the onset of the SN explosion. The binary separations are typically around three times the companion's radius.
In addition to the standard SD scenario, we consider extra systems that may have exited Roche-lobe overflow prior to explosion, for instance, due to prior mass transfer, a common-envelope phase, or spin-up/spin-down delays. For this reason, the binary separations in our models are fixed at approximately $1.5-3 R_\odot$, corresponding to $3-7 \times R_c$, which intentionally extend beyond the canonical Roche-lobe overflow range.
%In this study, the binary separations are fixed at approximately $1.5-3 R_\odot$, corresponding to $3-6 %\times R_c$. 
%which falls within the range of the standard SDS.

\begin{deluxetable*}{cccccccccc}
\tablewidth{0pt} 
%\tablenum{1}
\tablecaption{Simulations \label{tab:models}}

\tablehead{
Model Name & $M$ & $R$ & $A$ & $M_{\rm f}$ & $v_{\rm kick}$ & $v_{\rm orb}$ & $v_{\rm linear}$ & $M_{\rm b, Ni}$ & $L_{\rm f}$\\
 & [$M_\odot$] & [$R_\odot$] & [$R_\odot$] & [$M_\odot$] & [km~s$^{-1}$] & [km~s$^{-1}$] & [km~s$^{-1}$] & [$M_\odot$] & [g cm$^2$~s$^{-1}$]
}
\startdata 
%\cline{1-4}
M05A15 & 0.5 & 0.396 & 1.5 & 0.197 & 151 & (154) & (216) & $2.4 \times 10^{-10}$ & -- \\
M05A20 & 0.5 & 0.396 & 2.0 & 0.437 & 103 & (133) & (169) & $3.4 \times 10^{-9}$ & --\\
M05A30 & 0.5 & 0.396 & 3.0 & 0.492 & 45.5 & (109) & (118) & $8.6 \times 10^{-7}$ & --\\
M06A15 & 0.6 & 0.454 & 1.5 & $<0.07$ & -- & --  & -- & -- & -- \\
M06A17 & 0.6 & 0.454 & 1.7 & 0.315   & 142 & (137) & (198) & $5.1 \times 10^{-12}$ & -- \\
M06A18 & 0.6 & 0.454 & 1.8 & 0.372   & 133 & (133) & (189) & $1.8 \times 10^{-11}$ & --\\
M06A20 & 0.6 & 0.454 & 2.0 & 0.455   & 117 & (126) & (172) & $5.9 \times 10^{-10}$ & --\\
M06A30 & 0.6 & 0.454 & 3.0 & 0.574   & 56.8 & (103) & (118) & $6.4 \times 10^{-8}$ & --\\
M07A20 & 0.7 & 0.483 & 2.0 & 0.503   & 117 & (120) & (168) & $3.8 \times 10^{-10}$ & --\\
M07A30 & 0.7 & 0.483 & 3.0 & 0.654   & 61.4 & (98.3) & (116) & $3.2 \times 10^{-6}$ & --\\
\hline
M05A20-3D & 0.5 & 0.396 & 2.0 & 0.431 & 101 & 133 & 153 & $6.9 \times 10^{-7}$ & $6.55 \times 10^{49}$\\
M05A30-3D & 0.5 & 0.396 & 3.0 & 0.489 & 43.9 & 109 & 101 & $1.9 \times 10^{-5}$ & $5.20 \times 10^{49}$\\
M06A20-3D & 0.6 & 0.454 & 2.0 & 0.438 & 118 & 126 & 176 & $2.8 \times 10^{-7}$ & $6.63 \times 10^{49}$\\
M07A20-3D & 0.7 & 0.483 & 2.0 & 0.487 & 121 & 120 & 176 & $2.0 \times 10^{-7}$ & $7.84 \times 10^{49}$\\
M05A20-3D-noR & 0.5 & 0.396 & 2.0 & 0.436 & 99.0 & (133) & (166) & $9.0 \times 10^{-7}$ & --\\
M06A20-3D-noR & 0.6 & 0.454 & 2.0 & 0.451 & 113 & (126) & (170) & $3.6 \times 10^{-7}$ & --\\
M07A20-3D-noR & 0.7 & 0.483 & 2.0 & 0.501 & 114 & (120) & (166) & $2.6 \times 10^{-7}$ & --\\
\enddata
\tablenotetext{}{{\bf Notes.} $M$ is the mass of the companion at the onset of the SN explosion, $R$ is the corresponding stellar radius, $A$ is the binary separation, $M_{\rm f}$ is the final companion mass at the end of the hydrodynamic simulations, $v_{\rm kick}$ is the kick velocity due to the SN impact, $v_{\rm orb}$ is the orbital velocity at the onset of the SN explosion, $v_{\rm linear}$ is the final total linear velocity at the end of the hydrodynamic simulations, $M_{\rm b, Ni}$ is the final bound nickel mass, and $L_{\rm f}$ is the spin angular momentum of the surviving companion at the late stage of the hydrodynamic simulations. Model names ending with ``3D'' indicate three-dimensional simulations; otherwise, they are two-dimensional models. Model names ending with ``noR'' refer to 3D simulations performed without spin and orbital motions (i.e. without rotation). }
\end{deluxetable*}

%
% Figure 2
%

\begin{figure*}
\epsscale{0.55}
\plotone{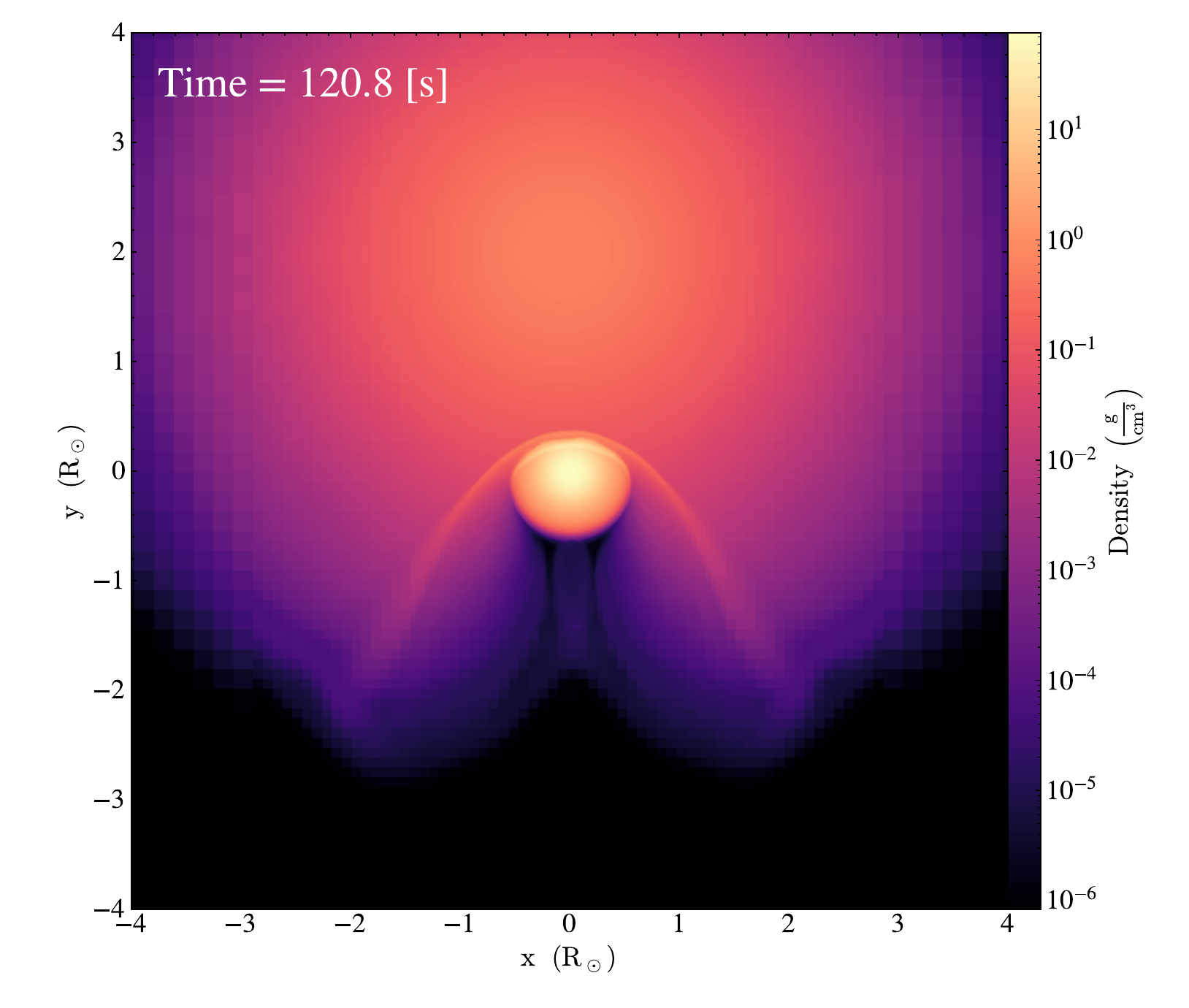}
\plotone{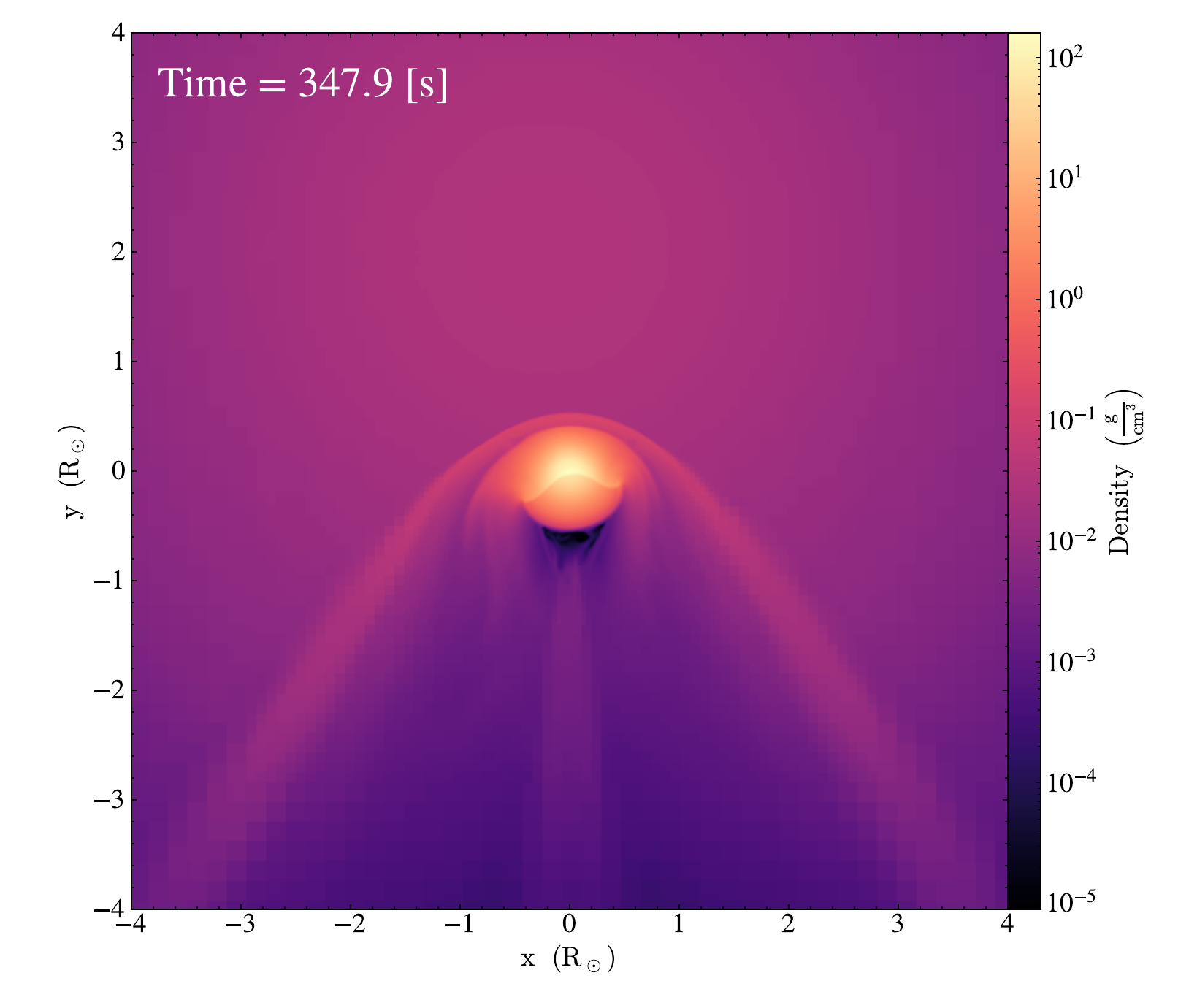}
\plotone{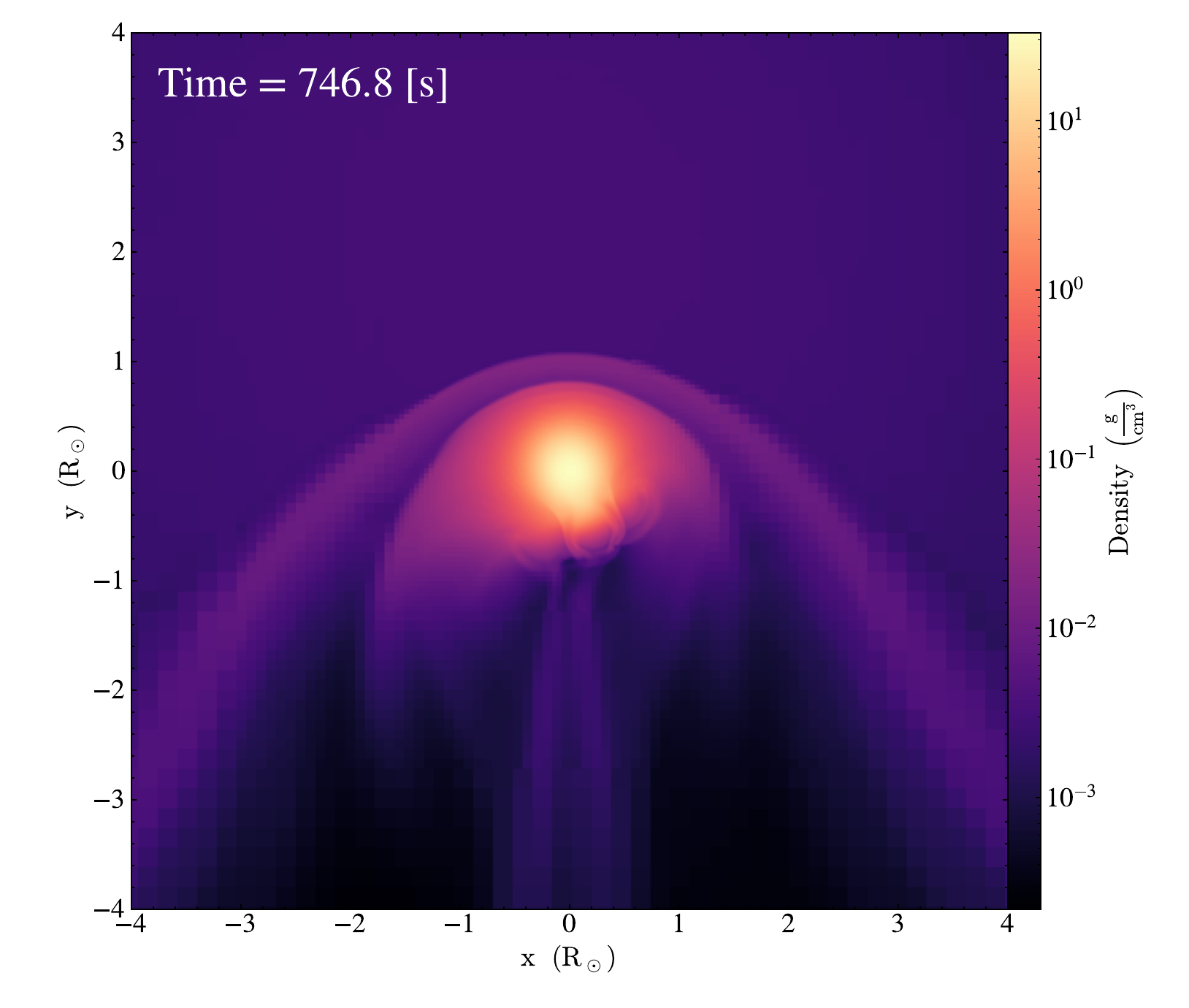}
\plotone{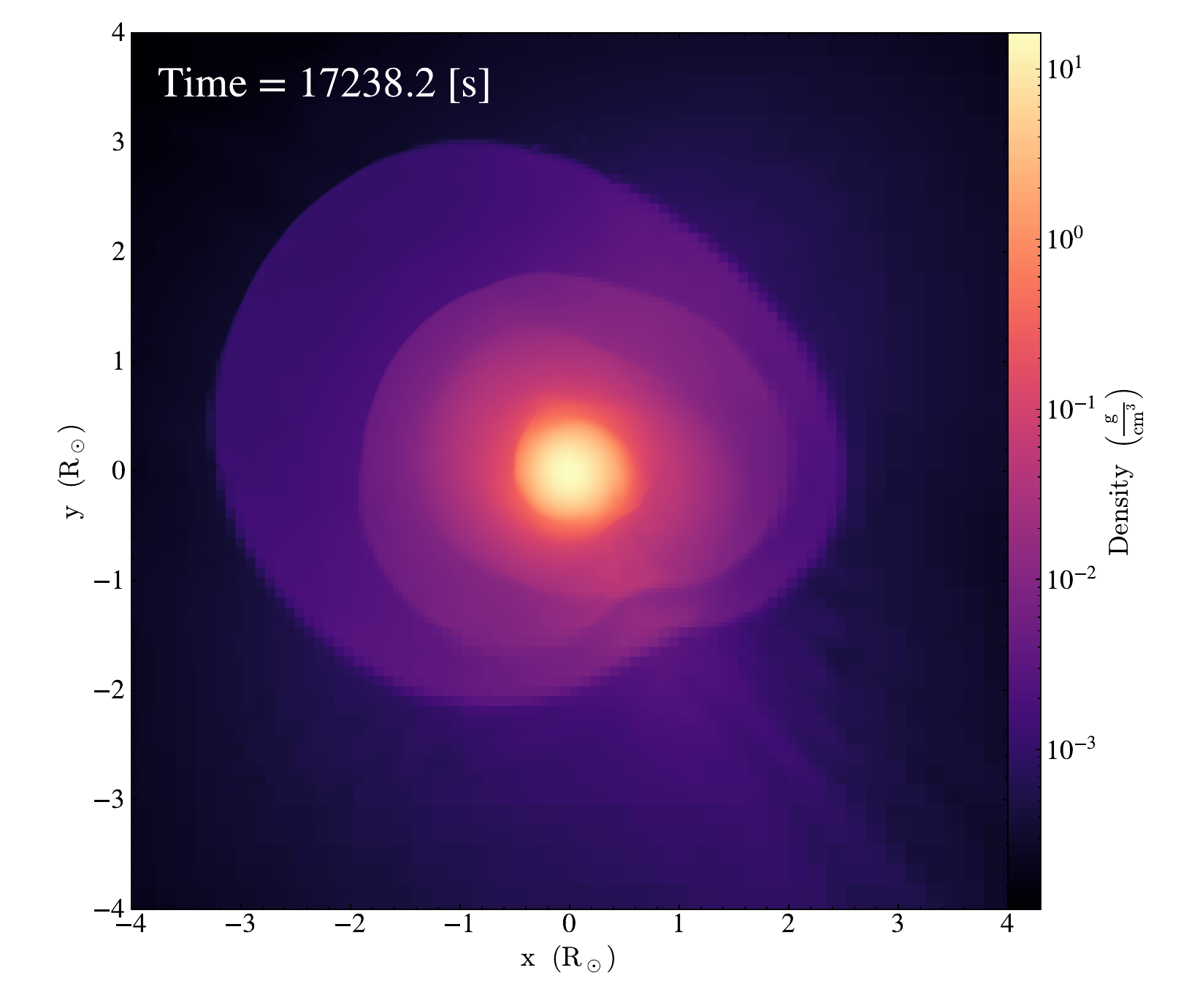}
\caption{Density distribution in the orbital plane at different time snapshots of model M06A20.
Each panel is re-centered on the center of the companion star.}
\label{fig:snimpact_M06A20}
\end{figure*}

\subsection{Surviving Companions}

To search for surviving companions in SN remnants, we would need to extend the simulations to cover hundreds of years. However, this is not possible with our hydrodynamics simulations, where the time step, on the order of seconds, is limited by the sound speed, and nuclear burning is ignored. 
Earlier attempts include using an ad-hoc toy model with artificial heating \citep{2003astro.ph..3660P, 2013ApJ...765..150S, 2015ApJ...805..170H}, reconstructing hydrostatic equilibrium models of post-impact companions \citep{2012ApJ...760...21P, 2013ApJ...773...49P}, or evaluating heating rates based on specific entropy changes \citep{2019ApJ...887...68B, 2022ApJ...933...38R, 2024ApJ...973...65W}.
For low-mass companions, we notice that during the SN impact, the companion star can be significantly compressed, with a total mass loss reaching as much as $50\%$. As a result, tracking the specific entropy changes after SN explosion becomes challenging in hydrodynamics simulations using Eulerian coordinates. Therefore, the entropy method should not be used in cases involving low-mass companions.

Thus, in this study, we adopt a simple analytical heating formula similar to the artificial heating equation used in \cite{2003astro.ph..3660P, 2013ApJ...765..150S, 2015ApJ...805..170H}. The heating rate is expressed as
\begin{equation}
\dot{\epsilon}(m) = \frac{\Delta E_{\rm heat}}{\tau_{heat} \sqrt{\pi \sigma} /2} \exp \left( - \frac{(1-m)^2}{\sigma} \right), \label{eq:heating}
\end{equation}
where $\Delta E_{\rm heat}$ is the amount of SN heating applied to the surviving companion, $\tau_{\rm heat} = 10^{-3}$~yr is the heating timescale, $\sigma$ describes the depth of the SN heating, and $m$ is the normalized enclosed mass.
We use {\tt MESA} version r24.03.1 to conduct the post-impact evolution.
First, we relax the mass of the pre-explosion companion models to the final bound mass from the hydrodynamics simulations to mimic the effects of mass stripping and ablation. 
We then use this heating formula to run post-impact evolution with all 3D models in Table~\ref{tab:models} with rotation.

\subsection{Stellar rotation and magnetic braking}

To enable rotation, we extra relax the specific angular momentum of the surviving companion model based on the spherically averaged profiles obtained from the hydrodynamic simulations with \texttt{FLASH}. This relaxation is performed using the \texttt{relax\_angular\_momentum\_filename} option in \texttt{MESA} , and it is applied before SN heating.
Since the surviving companion experiences a strong perturbation due to the supernova ejecta impact, the stellar wind of the surviving companion is expected to increase. In addition to employing the standard ``Dutch" wind scheme with the scaling factor {\tt Dutch\_scaling\_factor}=0.8, we introduce an additional uniform mass-loss rate of \(\dot{M} = 10^{-9}\,M_\odot\,\mathrm{yr}^{-1}\) for our surviving companion models. 
Note that this extra mass loss from a uniform stellar wind is required to have noticeable effects on carrying away angular momentum due to magnetic braking, but it has little impact on the bolometric luminosity and effective temperature evolution.

The surface magnetic fields of typical inactive and slowly rotating M dwarf stars are usually about a few hundred gauss \citep{2007ApJ...656.1121R}, but can reach several thousands gauss in fast-rotating stars \citep{2009ApJ...692..538R, 2017NatAs...1E.184S}. It is also unclear whether the magnetic fields would be destroyed or amplified during the SN impact. In this study, we incorporate magnetic braking into our models, assuming a surface magnetic field strength of \(B = 0,\;500,\;5000,\;\text{or}\;10,000\,\mathrm{G}\). 
Note that the high magnetic fields, such as $B=500$ or 5000~G, have been invoked to explain the orbital decays of black hole X-ray binaries \citep{2014MNRAS.438L..21G,2017MNRAS.465L..15G}.
The magnetic braking implementation follows the simple angular momentum loss formula described in \cite{1967ApJ...148..217W}, as provided in the standard \texttt{test\_suites} of \texttt{MESA}. Magnetic braking is activated whenever a mass loss occurs during the simulation.
Finally, we investigate the differences in the evolution with respect to different values of $\Delta E_{\rm heat}$ and $B$.

% ===========================================
%
% Figure 3
%

\begin{figure*}
\epsscale{0.5}
\plotone{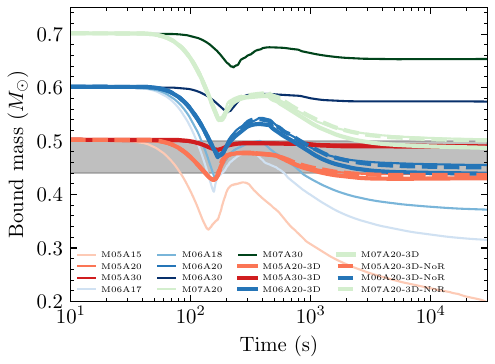}
\plotone{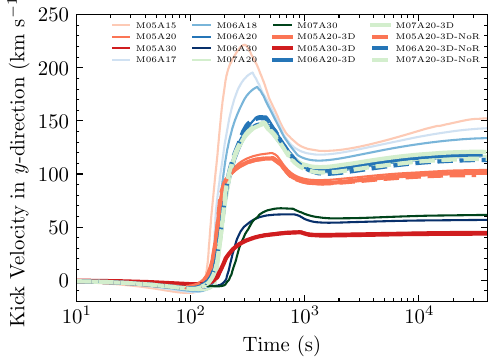}
\caption{{\it Left}: Evolution of the bound mass as a function of time. {\it Right}: Evolution of kick velocity (velocity along the direction perpendicular to the orbital velocity) as a function of time.
The gray band indicates the observed mass range of the surviving companion candidate MV-G272.
Different colors correspond to simulations with companion models of different initial masses. Transparency levels reflect the initial binary separations at the time of the SN explosion, with more transparent lines indicating shorter separations. Line thickness differentiates simulation dimensionality (2D or 3D). Dashed lines represent 3D simulations performed without spin and orbital motions ("noR" models). }
\label{fig:bmass}
\end{figure*}
% ===========================================

\section{Results} \label{sec:results}

\edit1{In this section}, we present the results of our multidimensional hydrodynamic simulations on the impact of supernova ejecta on a low-mass companion. We perform ten two-dimensional simulations using three low-mass companion models with masses of 0.5, 0.6, or 0.7 $M_\odot$. For each companion model, we explore different initial binary separations $A$. Table~\ref{tab:models} summarizes the parameter space explored in this study and lists a few important physical quantities from the simulation outcomes.
We select four representative cases from the two-dimensional simulations and conduct the corresponding three-dimensional simulations, including spin and orbital motions. In addition, we perform three additional three-dimensional simulations without spin and orbital motions as reference simulations (see Table~\ref{tab:models}).

In Section~\ref{sec:impact}, we describe the general evolution of the interaction between SN ejecta and its companion, and we present the final state of a surviving companion as observed at the end of the hydrodynamic simulations. In addition, we perform 64 post-impact {\tt MESA} simulations to investigate the effects of heating parameters (see Equation~\ref{eq:heating}) on the observable quantities in HR diagrams and to predict the linear and rotational velocities of these surviving companion models, as discussed in Section~\ref{sec:surviving}.

% ===========================================
\begin{figure*}
\epsscale{0.5}
\plotone{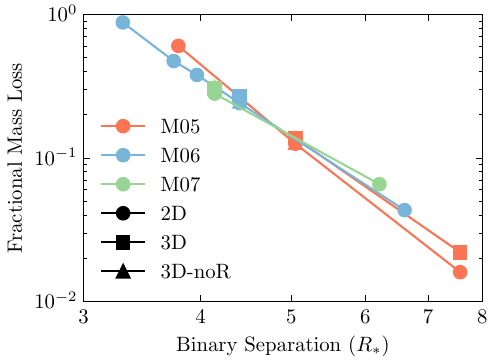}
\plotone{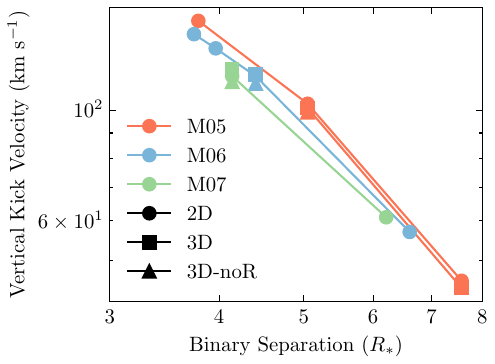}
\caption{{\it Left}: Fractional mass loss (unbound mass relative to initial mass) of the companion as a function of the initial binary separation (normalized by the companion radius). {\it Right}: Kick velocity perpendicular to the orbital motion as a function of the normalized binary separation. Different colors represent companion models with different initial masses. Markers differentiate simulation dimensionalities: 2D (circles), 3D-noR (triangles), and 3D (squares). }
\label{fig:powerlaw}
\end{figure*}
% ===========================================

\subsection{Impact of Supernova ejecta on a low-mass companion} \label{sec:impact}

The overall evolutionary features are similar to those of the MS companion models described in \cite{2010ApJ...715...78P, 2012ApJ...750..151P} and \cite{2022ApJ...933...38R}. However, the companions considered here are less compact than the $1-2 M_\odot$ MS models or He star models in those earlier works. For comparable separation-to-radius ratios, the companions experience a more intense impact and undergo greater mass loss due to their lower binding energies.

Figure~\ref{fig:snimpact_M06A20} shows the typical evolution of the density in the orbital plane at different times after the SN explosion for Model M06A20-3D. At approximately 100~s after the explosion, the supernova ejecta collide with the companion's near side, compressing the star and driving a strong bow shock around it
This initial impact begins to ablate the outer envelope of the companion. Within a few minutes post-impact ($t \sim 300-700$ s), the shock penetrates deeper into the star's interior, and the stripping of mass intensify dramatically. Although the ejecta speed still far exceeds the companion's orbital or spin velocity, asymmetric features appear on the back side of the companion due to the influence of its spin and orbital motion.

At later times (on the order of an hour and beyond), after the main shock front has passed, the surviving remnant of the companion, now substantially lighter ($M_{\rm f}=0.455 M_\odot$, see Table~\ref{tab:models}), expands and undergoes radial oscillations as it seeks a new equilibrium state.
In addition, momentum transferred from the supernova ejecta propels the surviving companion outward from the explosion center, imparting a kick velocity that transforms the star into an unbound runaway, carrying both its original orbital motion and an extra velocity component from the explosion itself.

Figure~\ref{fig:bmass} shows the evolution of the bound mass as a function of time for all hydrodynamic simulations. The bound mass is defined by the total mass of zones with negative total energy. 
The bound mass of the companion drops rapidly as the SN ejecta strips its envelope, with the more severe mass loss occurring for tighter binaries and more massive companions.
For example, model M06A15, a 0.6 $M_{\odot}$ companion at $A=1.5,R_{\odot}$, is almost completely ablated ($M_{f}<0.07,M_{\odot}$), whereas at a wider separation of $A=3.0,R_{\odot}$ (model M06A30) it retains over 95\% of its mass ($M_{f}\approx0.57,M_{\odot}$).
At an intermediate separation ($A=2.0,R_{\odot}$), the final bound masses are $M_{f}\approx0.44,M_{\odot}$, $0.46,M_{\odot}$, and $0.50,M_{\odot}$ for companions of initial mass 0.5, 0.6, and 0.7 $M_{\odot}$, respectively (see Table~\ref{tab:models}).
Notably, these values lie in the $\sim0.44$–$0.50,M_{\odot}$ range marked by the gray band, consistent with the observed mass of the MV-G272 companion. 

We also find excellent agreement between the 2D and 3D-noR simulations, which yield nearly identical final bound masses. The 3D models that include the companion’s rotation (and orbital motion) exhibit only a modest additional mass loss, appearing mainly at late times as the rotating envelope sheds a bit more material. As a result, their final $M_{f}$ values remain within a few percent of the non-rotating cases, indicating that rotation induces only a slight enhancement in mass stripping.

The right panel of Figure~\ref{fig:bmass} shows the evolution of the kick velocity imparted to the surviving companion, specifically the $y$-component perpendicular to its orbital motion. This choice of velocity highlights the kick velocity and makes it easier to compare simulations with or without rotation. 
The kick accelerates the star mostly in the radial direction (away from the SN) and saturates once the ejecta passage is complete.
As expected, closer companions receive larger kicks: for a 0.5 $M_{\odot}$ star, $v_{\rm kick}\approx150$ km~s$^{-1}$ at $A=1.5, R_{\odot}$, versus $\approx45$ km~s$^{-1}$ at $A=3.0, R_{\odot}$.
We also see that the 2D and 3D-noR models produce virtually the same $v_{\rm kick}(t)$ evolution, implying that asymmetric 3D effects are negligible in the absence of spin. Only when the companion’s spin is included (3D models) does a slight excess kick emerge at late times, adding on the order of a few km~s$^{-1}$ to the final velocity.
Combining the kick (in $y$) with the pre-SN orbital velocity (tangential component) gives a resultant runaway speed of order $150$–$200$ km~s$^{-1}$ for the ejected companion in these simulations (see $v_{\rm orb}$ and $v_{\rm linear}$ in Table~\ref{tab:models}).

Figure~\ref{fig:powerlaw} demonstrates that both the fraction of companion mass lost and the imparted kick velocity scale strongly with the pre-supernova orbital separation, following clear power-law trends.
This trend is consistent with previous finding, such as \cite{2000ApJS..128..615M, 2012ApJ...760...21P, 2012A&A...548A...2L, 2013ApJ...773...49P, 2019ApJ...887...68B, 2022ApJ...933...38R, 2024ApJ...973...65W}.
Quantitatively, excluding the closest-separation outlier (Model M05A15, which undergoes near-total disruption), the simulation data are well-fit by power-laws: the fractional mass loss roughly scales as $a^{-4.5}$ (a power-law index of order 3), while the kick velocity scales approximately as $a^{-1.6}$. 
These power-law relations provide a valuable interpretation when degeneracy arises between the final mass of a surviving companion and its initial binary separation.

% ===========================================
\begin{figure}
\epsscale{1.2}
\plotone{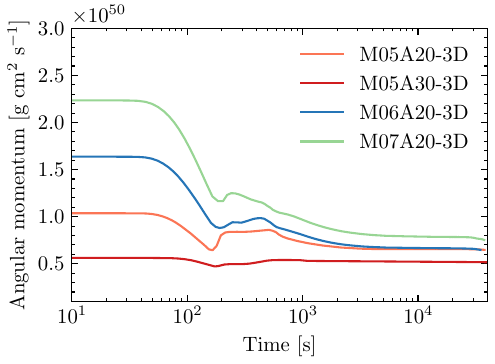}
\caption{The evolution of the spin angular momentum magnitude as a function of time is shown. Different colors represent the angular momentum magnitudes from various 3D models with pre-explosion spin and rotation. The spin angular momentum is computed in the center-of-mass frame of the surviving companion, considering only the bound gas.}
\label{fig:angularmomentum}
\end{figure}
% ===========================================

\subsection{Angular momentum loss}

Before the SN explosion, the companion star is expected to be a fast rotator due to tidal locking in the close binary. In our 3D impact simulations, we initialize the companion with a spin-to-orbital-period ratio of 0.95 (almost synchronous rotation). This means the pre-explosion companion has a high spin angular momentum, as it nearly co-rotates rapidly with the orbit. We track how this spin evolves once the SN ejecta strikes the star.

Figure~\ref{fig:angularmomentum} shows the evolution of the companion’s spin angular momentum (magnitude) as a function of time for several representative 3D models that include the initial rotation. We calculate the spin angular momentum of the bound remnant (excluding unbound ejecta) in the companion’s center-of-mass frame. It is evident that the SN impact causes a net loss of angular momentum from the star. As the supernova ejecta sweeps over the companion, it strips away a portion of the star’s outer layers and, with them, carries off a substantial fraction of the star’s angular momentum. This is a key channel for the spin-down of a surviving companion as reported in \cite{2012ApJ...750..151P, 2013A&A...554A.109L}. 

\cite{2012ApJ...750..151P} found an angular momentum loss of about 50\% for a 1.17 $M_\odot$ MS companion model.
However, the degree of angular momentum loss varies significantly between models in this study. 
As shown in Figure~\ref{fig:angularmomentum}, some models exhibit a much steeper decline than others.
For instance, in the tight-separation, higher-mass case (model M07A20-3D), the companion experiences substantial angular momentum loss during the SN impact, retaining only $\sim 30$\% of its initial spin angular momentum by the end of the interaction. In contrast, the wider-separation, lower-mass model (M05A30-3D) shows very little change, preserving most of its original spin.
Therefore, some of our models require additional mechanisms for angular momentum loss to explain the slow rotation observed in surviving companion candidates.  

\begin{figure}
\epsscale{1.2}
\plotone{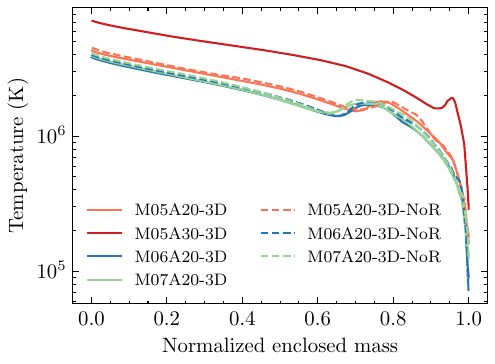}
\caption{Temperature profiles of the surviving companion models at the end of the hydrodynamic simulations. Different colors correspond to simulations with different companion masses. Solid and dashed lines distinguish simulations performed with and without rotation (spin and orbital motions), respectively.}
\label{fig:temperature_profiles}
\end{figure}

\begin{figure*}
\epsscale{0.9}
\plotone{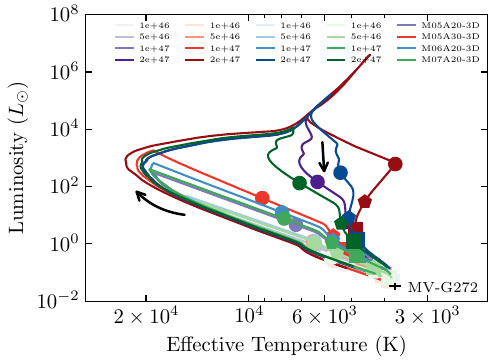}
\plotone{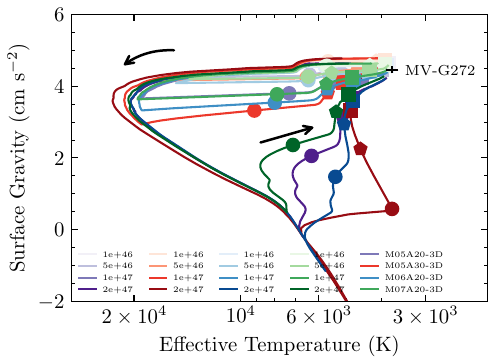}
\caption{Evolution of the surviving companion models in the Hertzsprung–Russell (HR) diagram. Different colors represent different 3D companion models with different magnitudes of SN heating ($\Delta E_{\rm heat}$). Symbols mark evolutionary stages at different times after the SN explosion: circles, pentagons, and squares represent their locations at $t= 0.1, 100$, and 7500 years, respectively. Magnetic fields are set to zero in these models.}
\label{fig:HR}
\end{figure*}

\begin{figure*}
\epsscale{1.2}
\plotone{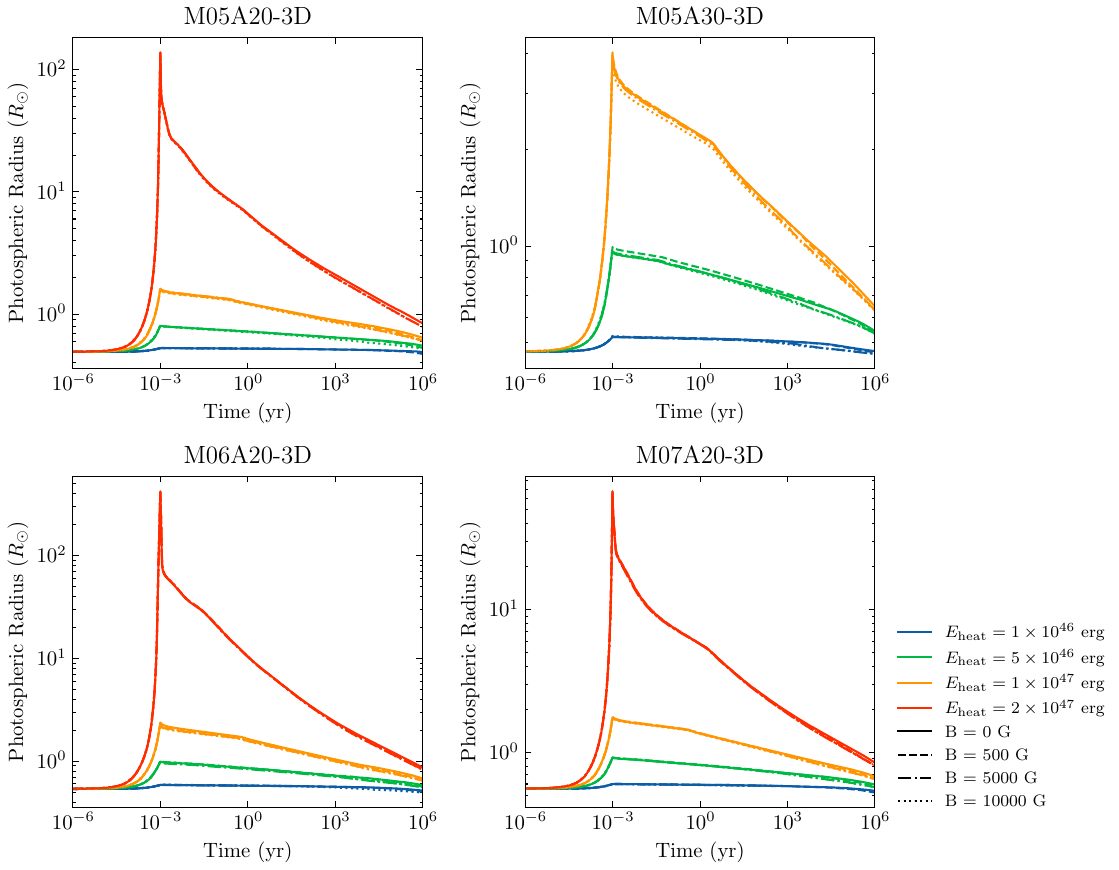}

\caption{Time evolution of the photospheric radius of surviving companion models after the SN impact. Different panels represent different initial companion models, and line styles correspond to simulations with different assumed surface magnetic field strengths. }
\label{fig:surv_radius}
\end{figure*}

\begin{figure*}
\epsscale{1.2}
\plotone{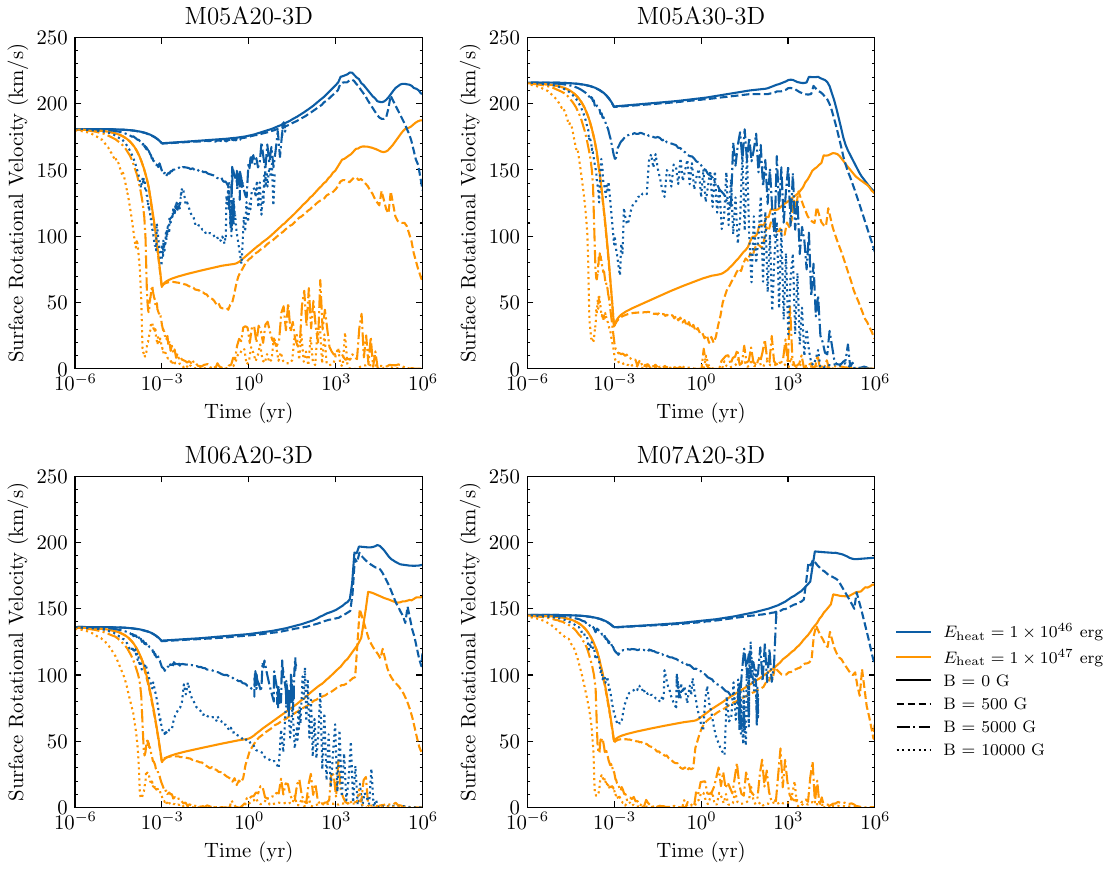}

\caption{Evolution of the surface rotation velocity of the surviving companion models after the SN impact. Different panels correspond to different initial companion models. Line styles represent simulations with various assumed magnetic field strengths.
We only show models with $E_{\rm heat} = 10^{46}$ erg and $10^{47}$ erg for better clarity.}
\label{fig:surv_rot}
\end{figure*}

\subsection{Post-impact evolution} \label{sec:surviving}

An important factor governing the post-supernova evolution of the surviving companion is the depth at which the ejecta’s energy is deposited in the star. 
This depth effectively sets the local thermal diffusion timescale for the release of the deposited energy. Energy deposited deeper in the envelope will take longer to diffuse out, prolonging the star’s response, whereas shallow deposition leads to a faster release. 
In our toy model, we control the heating depth using a dimensionless parameter, $\sigma$ (defined in Equation~\ref{eq:heating}), which represents the fraction of the stellar mass (or radius) over which the SN impact energy is injected. 
We calibrate $\sigma$ by comparing it to the detailed hydrodynamic impact simulations. 
Figure~\ref{fig:temperature_profiles} shows the interior temperature profiles of several companion models at the end of the 3D impact simulations. 
By matching these profiles, we infer that a relatively shallow heating ($\sigma \sim 0.2$) best reproduces the post-impact temperature structure in the tight-separation cases (A20 models), 
whereas a much smaller value ($\sigma \sim 0.05$) is more appropriate for the wider-separation case (A30). 
This reflects the fact that in closer binaries, the SN shock heats a larger fraction of the companion’s envelope, while in wider binaries, the energy deposition is confined to a thinner outer layer. 

We estimate the amount of energy injected into the companion’s interior using the geometric interception of the SN explosion energy. If $E_{\mathrm{SN}}$ is the total ejecta kinetic energy, the companion of radius $R_c$ at a separation $A$ would intercept roughly a fraction $f \sim \frac{R_c^2}{4A^2}$ of the explosion (assuming isotropic ejecta).
Not all of this intercepted energy is absorbed as heat—indeed, the majority goes into unbinding and blowing off the outer layers of the companion’s envelope. We therefore introduce an efficiency factor $\xi$ (on the order of $10^{-2}$–$10^{-1}$) to account for the fraction of intercepted energy that is actually deposited as thermal energy in the star. This gives a rough heating budget of
\begin{equation}
\Delta E_{\rm heat} = \xi E_{\rm SN} \times \left( \frac{R_c^2}{4A^2} \right).
\end{equation}  
We consider the range of the $\xi = 10^{-1} - 10^{-2}$, corresponding to $\Delta E_{\rm heat} = 10^{46} - 10^{47}$~erg. 
Guided by this estimate, we perform a suite of post-impact evolution simulations with four representative energy injection values: $1\times10^{46}$, $5\times10^{46}$, $1\times10^{47}$, and $2\times10^{47}$ erg. These values span the expected range of thermal energy deposited in the companion by the SN impact.

The evolutionary tracks of the heated companions in the Hertzsprung–Russell diagram are shown in Figure~\ref{fig:HR} (top panel), plotting effective temperature ($T_{\rm eff}$) against bolometric luminosity. All models start near the lower-right region of the diagram (cool $T_{\rm eff}$ and low $L_{\rm bol}$, characteristic of the pre-impact companion) and then loop clockwise through the HR diagram. Within the first $\sim$1~year after the impact, the star brightens substantially, moving upward and slightly toward the right (cooler $T_{\rm eff}$) in the diagram. This initial brightening is powered directly by the release of the deposited shock energy, which causes the envelope to heat up and expand. The expansion of the outer layers leads to a larger radius and a momentary drop in surface temperature, hence the rightward drift despite the rising luminosity. As the injected energy diffuses outward and is radiated away, the star reverses course in the HR diagram: the photosphere begins to contract back inward. During this contraction phase, gravitational potential energy is released, which helps sustain the luminosity even as the star’s total thermal energy content diminishes. The track consequently turns back toward higher $T_{\rm eff}$ (leftward) while gradually fading in $L_{\rm bol}$, completing a clockwise loop. The lower panel of Figure~\ref{fig:HR}  shows the corresponding evolution of $T_{\rm eff}$ vs. surface gravity ($\log g$). In this representation, the impact of the transient inflation is evident as a drop in $\log g$ (when the star’s radius is near its peak) followed by a rise in $\log g$ as the star contracts back toward its original compact state.

Figure~\ref{fig:surv_radius} illustrates the radius evolution of the post-impact companion, highlighting the effects of different assumed surface magnetic field strengths (different line styles). In all cases, the overall radial evolution is qualitatively similar: the star undergoes a rapid expansion immediately after energy deposition, then gradually contracts over the following years to decades as it thermally relaxes. We find that including magnetic braking (for the range of field strengths considered) has very little influence on the radius vs. time behavior. Even with a strong magnetic field, the pressure-driven expansion and subsequent cooling contraction of the star proceed almost unchanged. Consequently, the presence of a moderate magnetized wind does not appreciably alter the star’s path in the HR diagram or its luminosity/temperature evolution. We note that as the companion’s radius decreases at late times, conservation of angular momentum would tend to spin up the star’s surface rotation in the absence of external torques. In other words, the contraction itself causes an increase in the spin rate, a point we examine in detail with the inclusion of magnetic braking in our rotation models.

Figure~\ref{fig:surv_rot} shows the evolution of the companion’s surface rotation velocity under the influence of magnetic braking for various assumed magnetic field strengths. As expected, higher initial field strengths lead to more efficient angular momentum loss via the stellar wind and, thus, a faster spin-down of the star. In the absence of magnetic braking, the post-impact contraction would cause the surface rotation velocity ($v_{\rm rot}$) to rise (due to the shrinking moment of inertia). However, with sufficiently strong magnetic torques, this spin-up can be counteracted and reversed. Our simulations indicate that extraordinarily strong surface fields on the order of $B \gtrsim 5$ kG (kilogauss) are required to reduce the rotation to $v_{\rm rot} < 10$ km s$^{-1}$ within a practical observational timeframe after the explosion. For example, with $B \approx 5$ kG, the companion’s rotation is efficiently braked to a slow spin by the time the SN remnant is a few thousand years old. We have assumed a constant stellar wind mass-loss rate of $10^{-9},M_{\odot}$ yr$^{-1}$ in these post-impact evolution calculations, consistent with a robust magnetized wind from a late-type star. We note that for weaker field strengths (significantly below a few kG), the magnetic braking is too feeble to offset the contraction-induced spin-up, and the survivor remains a relatively fast rotator. Thus, while the hydrodynamic impact itself removes a large fraction of the companion’s original angular momentum (Section 4.2), an additional mechanism—such as a very strong magnetic wind—is likely necessary to brake the star’s rotation to the low values that might be observed in candidate SN companion stars as suggested by \cite{2023ApJ...947...90R}.

\section{Discussions}

%The observed quantities of G272.2-3.2 are taken from \cite{2023ApJ...947...90R}, where we consider $L=10^{-1.39} L_\odot$, $T_{\rm eff} = 3800$~K, and $\log g = 4.46$~cm~s$^{-2}$. We also assume the age of the supernova remnants is $\sim 7500$ yrs.

In this section, we discuss the implications of our numerical results in the broader astrophysical context. In Section~\ref{sec:vlinear}, we focus on the spatial linear velocities of surviving M-dwarf companions, highlighting how these velocities can serve as observational signatures to identify potential survivors within supernova remnants and comparing our simulation predictions with known observational cases. Section~\ref{sec:rotation} addresses the unexpectedly slow rotation observed in candidate companion MV-G272. In particular, we focus on the role of magnetic braking as a mechanism for angular momentum loss. Finally, we show that M dwarf surviving companions might has very little surface aboundance abnormal due to the SN ejecta contamination in Section~\ref{sec:contamination}.

\subsection{Spacial velocity of surviving companions \label{sec:vlinear}}

A number of high-velocity stars have been proposed as companions of SNe Ia produced in 
the {\it dynamically-driven double-degenerate double-detonation} ($D^{6}$) scenario of \cite{2018ApJ...865...15S}. 
Three hypervelocity white dwarfs (D6-1, D6-2, D6-3) were found by \cite{2018ApJ...865...15S}, 
from their high proper motions shown in the {\it Gaia} DR2, hypervelocity meaning space velocities 
$v > 1000$ km s$^{-1}$. 
Six other runaways have been found by \cite{2023OJAp....6E..28E}, using the {\it Gaia} DR3. 
The hypervelocity white dwarfs are supposed to be the surviving mass-donors in a $D^{6}$ SN Ia. 
Other high-velocity white dwarfs (with $v < 1000$ km s$^{-1}$) could come from mass-accretors 
only partially burned and ejected as well from a close binary system that produced a SN of the 
SN Iax type \citep{2019MNRAS.489.1489R}.
In our simulations with M-dwarf companions emerge as runawys with total speeds on the order of 150-200~km$^{-1}$.
This velocity scale is somewhat lower than the typical hypervelocity stars predicted for D6 models,
but it is comparable to (or higher than) the velocities expected for more massive main-sequence or red-giant companions in single-degenerate scenarios \citep{2012ApJ...750..151P}.
In all cases, a surviving companion with substantial space velocity (see Table~\ref{tab:models}), moving a way from the SN explosion center is expected in our simulations.

The Gaia mission has made it feasible to detect such runaway companions by their proper motions. In the case of SNR G272.2-3.2, the M-dwarf star MV-G272 stood out as an $8.9\sigma$ outlier in proper motion relative to the local stellar sample \citep{2023ApJ...947...90R}. Its total proper motion of 38.15 mas~yr$^{-1}$, combined with a Gaia distance of $1.32^{+1}_{-0.39}$ kpc, implies a tangential velocity of order $\sim 200-240$~km~s$^{-1}$, 
yielding an estimated 3D space velocity around $\sim 256^{+181}_{-70}$~km~s$^{-1}$ \citep{2023ApJ...947...90R}. 
This is consistent with the velocities from our M-dwarf companion simulations. Notably, rewinding MV-G272’s motion by ~7500 years (the approximate age of the remnant) brings it near the geometric center of G272.2-3.2, whereas other stars with large proper motions in the field do not trace back to the remnant center. 
The high velocity and the trajectory pointing away from the explosion site make MV-G272 an excellent surviving companion candidate.
In addtion, the observed effective temperature ($T_{\rm eff}=3600-3850$) and surfface gravity ($\log g =4.36$) from the CARMENES VIS spectra \citep{2023ApJ...947...90R} are also consistent with our models with low $\Delta E_{\rm heat}$ (see the lower panel in Figure~\ref{fig:HR}).

However, there are practical limitations in using spacial velcoity as a sole identifier of surviving companions.
Proper motions give only the transverse component of a star's velocity, or the line-of-sight velocity will go unnoticed without spectra. In addition, if the surviving companion’s velocity vector is not perfectly radial with respect to the remnant (for example, if the binary had a significant center-of-mass motion or the kick was asymmetrical), the star’s path may not cleanly intersect the remnant’s geometric center. This necessitates careful analysis of the full 3D space motion when possible.

Another difficulty is that fast-moving companions can quickly escape the vicinity of the remnant.
In an ancient remnant of order $10^5$~years, 
even a modest 200~km~s$^{-1}$ runaway could by now be $\sim 20-30$~pc away, 
well outside the faint remnant debris. 
This may explain why searches in historical SN~Ia remnants have often come up empty. In addition, the definition of a SNR center has its own uncertainity as well. 
Therefore, additional evidence such as the rotation of a surviving companion could place additional constraint on the progenitor systems. 

\subsection{The rotation of MV-G272 \label{sec:rotation}}

Beyond high spacial velocity, a surviving companion may also bear the imprint of the supernova on its rotation \citep{2012ApJ...760...21P, 2013A&A...554A.109L}. 
Prior to the explosion, the companion in a close binary would have been tidally locked, spinning at the same rate as its orbit.
However, observationaly, there is no information about rotation but for one hypervelocity and one high-velocity white
dwarf, assumed to come from one of these two scenarios each: LP 398-9 (D6-2) (hypervelocity: \citealt{2022MNRAS.512.6122C}) 
and LP 40-365 (GD 492) (high-velocity, partially burned: \citealt{2021ApJ...914L...3H}). 

In the two cases, the white dwarfs are slow rotators, in spite of coming from systems with 
high orbital velocities.
LP 393-9 has a rotation period of P = 15.4 hr. Its radius being $R = 0.20 \pm 0.01 R_{\odot}$, 
that means a rotational velocity of $v_{\rm rot} = 80 \pm 10$ km s$^{-1}$, to be compared with
a space velocity of $v = 1013 \pm 61$ km s$^{-1}$, that is, $v_{\rm rot}$ is only some 8\% of
$v$. In the case of LP 40-365, with a period $P = 8.914$ hr and radius $R = 0.16 \pm 0.01 R_{\odot}$, (Raddi et al. 2019) we have $v_{\rm rot} = 15.776$ km s$^{-1}$, in front of 
$v \simeq 837$ km s$^{-1}$, so $v_{\rm rot}$ is a mere $\simeq$ 1.9\% of $v$.  In the case of 
MV-G272, we have that $v_{\rm rot}$ is $\simeq$ 1.1\% of $v$, so within the range of these other 
two proposed SN Ia surviving companions (see Table~\ref{tab:observations}).
\cite{2022MNRAS.512.6122C}, to explain the rotation velocity of LP 398-9 invoke the possible role 
of magneting braking.

Star MV-G272 is also slow rotator, its measured rotational velocity being $v_{\rm rot} \sin i < 4$ km s$^{-1}$ \citep{2023ApJ...947...90R}.
In our hydrodynamics simulations, as shown in Figure~\ref{fig:angularmomentum}, 
the impact of the SN ejecta could reduce a certain amount of the angular momentum (also the surface rotation), depending on the models. 
Removal of material from the companion takes place by two different mechanisms:
on impact, the layers closer to the surface gain enough momentum to immediately
leave the star (``stripped material''), while deeper layers just absorb enough
thermal energy to become unboud and leave the star as a strong wind (``ablated'' material). 

However, even a significant amount of angular momentum is lost during the supernova impact in Model M07A20-3D, a fast surface rotation ($> 130$~km~s$^{-1}$) could still be there after the SN impact if there is no magnetic braking (see Figure~\ref{fig:surv_rot}). 
If M dwarf stars possess significant magnetic fields ($>5$~kG), so
that ``ablated'' material can remain linked to the surface and co-rotate
with it up to some distance, from being permeated by the fields. Magnetic
braking of the rotation can thus take place.

In the numerical simulations above, all the material lost by the companion
upon impact of the SN ejecta does not further interact with it after
becoming separated from the new surface. Interaction via the surface magnetic
field has not been considered.
Full magnetohydrodynamical modeling is required to properly address the
question.

\begin{figure}
\epsscale{1.2}
\plotone{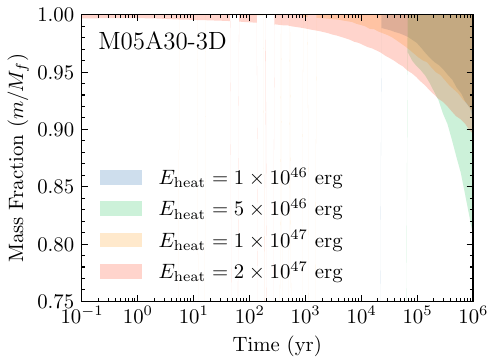}

\caption{Convective mixing region in Model M05A30. The evolution of the mass fraction of the convectively mixed region is shown over time for models with varying supernova heating energies. The magnetic field B is set to zero in all these simulations.}

\label{fig:mx1M05A30}
\end{figure}

\subsection{Surface contamination \label{sec:contamination}}

Another observational signature of a surviving companion in a SN Ia event is the potential surface contamination resulting from the supernova ejecta. Following the explosion, a small amount of supernova material can become gravitationally bound to the surface of the surviving companion. This contamination may lead to enhanced abundances of heavy elements, such as nickel or iron, observable in the star's spectral lines. 
\cite{2010ApJ...715...78P, 2012ApJ...750..151P} suggest the amount of nickel contamination could be as high as $5 \times 10^{-4} M_\odot$ for a helium star companion or approximately $10^{-5} M_\odot$ for sun-like main-sequence companions. In our models, as presented in Table~\ref{tab:models}, the nickel contamination in M dwarf companions is typically around $10^{-7} M_\odot$. Even in the most extreme case, model M05A30-3D, the amount of bound nickel is only about $1.9 \times 10^{-5} M_\odot$.

Observationally, detecting such contamination has proven challenging. Searches for surviving companions in historical SN Ia remnants, such as Tycho's supernova \citep{2009ApJ...691....1G, 2013ApJ...774...99K, 2014MNRAS.439..354B} and SN 1006 \citep{2012Natur.489..533G}, have not yielded definitive evidence of surface contamination.
The absence of observational evidence for surface contamination in surviving companions suggests several possibilities. It is possible that the amount of accreted material is below current detection thresholds or that the contaminants have been mixed into deeper layers of the star, making them less observable.

Figure~\ref{fig:mx1M05A30} shows the evolution of the convective region in our M05A30 models. For the most strongly heated case, the convective envelope extends from the surface inward to about 97\% of the stellar mass by 7500 years after the explosion, corresponding to a convective zone mass of $0.015M_\odot$ (i.e. 3\% of a 0.5 $M_\odot$ M dwarf). 
Assuming full mixing within this zone and that all of the bound nickel ($2\times 10^{-5} M_\odot$) decays to iron, we can estimate a surface iron mass fraction $X_{\rm Fe, new}= X_{\rm Fe, SN} + X_{\rm Fe, \odot} \sim 2 \times 10^{-5}/ 0.015 + X_{\rm Fe, \odot} \sim 2.3 \times 10^{-3}$, compared to the solar value $X_{\rm Fe, \odot} \sim 10^{-3}$. This yields an iron abundance enhancement of [Fe/H]$= \log_{10} \left(\frac{X_{\rm Fe, new}}{X_{\rm Fe, \odot}} \right) = 0.36$. 
Since no additional nickel is added, the nickel-to-iron ratio becomes diluted, leading to [Ni/Fe]$=-0.36$. These values suggest that while the total heavy element contamination may be modest, detectable abundance anomalies could still persist in the atmospheres of surviving M dwarf companions, particularly in iron lines, depending on mixing efficiency and observational sensitivity.

%Mconv = 0.97 at 7500 years, give $M_{\rm conv} = 0.03 \times 0.5 = 0.015M_\odot$.
%Assume all nickel decay to iron (no additional Ni is added).
%Assume solar abundance: Fe = 1e-3, Ni = 8e-5
%Solar Fe mass in convective envelope = $10^{-3} \times 1.5 \times 10^{-2} = 1.5 \times %10^{-5} M_\odot$.
%Fe mass with SN contamination = Fe solar + Fe bound from nickel = 1.5e-5 + 2e-5 = $3.5 \times 10^{-5} M_\odot$.
%New Fe mass fraction = (3.5e-5 / 1.5e-2) = 2.3e-3
%Thus $[Fe/H] = \log_{10} \left(\frac{2.3 \times 10^{-3}}{10^{-3}} \right)$ = 0.36 
%Ni mass stays the same as solar = 8e-5
%New Ni/Fe ratio = 8e-5 /2.3e-3 = 3.5e-2
%solar Ni/Fe ratio = 8e-2
%Thus $[Ni/Fe] = \log_{10} \left(\frac{3.5 \times 10^{-2}}{8 \times 10^{-2}} \right)$ = -0.36 

\section{Summary \& Conclusions}

The problem of the nature of the stellar systems giving rise to SNe Ia remains open after many years of observations and theoretical modelling.
There is no complete certainty in favour of any of the proposed progenitor systems nor explosion mechanisms. Searches in the remnants of historical Galactic SNe Ia (SN 1006, Tycho and Kepler SNe) have just concluded in the absence of surviving SNe Ia companions in SN 1006 \citep{2012Natur.489..533G} and Kepler \citep{2018ApJ...862..124R}. See further search results in \cite{2019NewAR..8501523R} and references therein, and absence of surviving
SNe Ia companions  in the LMC such as in SNR 0509-67.5 \citep{2023ApJ...950....10L}.   

Exploration of a new, older ($\sim$ 7500 yr) Galactic SNR, G272.2-3.2, has found a candidate star, named MV-G272, strongly favoured by its kinematics.
It is a M1-M2 dwarf and is slowly rotating. That poses the problem of whether a star in its range of mass can possibly have the present luminosity and surface temperature of MV-G272 and not be rotating faster.
 
Pre-explosion models of M dwarf companions in the range of masses 0.5-0.7 $M_{\odot}$ have first been constructed using the MESA code. The impact of the ejecta of a SN Ia on these models has been simulated in 2D and 3D. Different separations between the exploding WD and its companion have been considered. The initial models, in the 3D simulations, had high spin angular momenta.

The effects of stripping, heating and angular momentum loss given by the hydrodynamic simulations have been incorporated in the initial post-impact models. Since there is strong compression and huge mass loss in the case of M-dwarfs, tracking the specific entropy changes has been replaced by an artificial heating equation.

Although there is considerable angular momentum loss at this stage, it does not yet bring the rotation down to the MV-G272 value. No magnetic fields have been included here, however.

The post-impact evolution has then been followed, including the effects of mass loss on the angular momentum of the star. A wind appropriated for a strongly perturbed M dwarf has been implemented and magnetic braking has been incorporated into the models, for several magnetic field strengths.

The thermal evolution of the surviving companion  depends on the depth at which the energy of the ejecta is deposited. Such depth is calibrated from the hydrodynamic simulations of the impact of the SN ejecta. It is found that a relatively shallow heating region best reproduces the post-impact temperature structure. That also depends on the separation of the companion from the exploding WD. Effective temperatures and luminosities like those of MV-G272 do result, for times corresponding to the age of the SNR.

The braking of the initial rotation of the companion by the immediate effect of the impact, plus the subsequent stellar wind coupled to the surface magnetic field, during the post-explosion evolution, can bring the rotation down to the MV-G272 value. Coupling of the material ablated in the explosion with the star's magneting field would add extra braking.

Surface contamination from SN material captured by the companion has been estimated in the form of nickel mass, that being small. Adding to this the diluting effect of convection, the result is entirely compatible with
the measured surface abundances of MV-G272.

In conclusion, an object with the mass, linear velocity, effective temperature, luminosity and spin of MV-G272 can be the surviving companion of the SN Ia that produced the G272.2-3,2 SNR. Its slow rotation can be explained
by the combined effects of mass striping and ablation at the time of explosion plus magneting braking during the post-explosion evolution.

Most of what has been learnt does not only concern MV-G272 but to other possible M star companions of low mass and at different separations from the SN Ia.

Observationally MV-G272 and the stars studied here would not be the only SNe Ia companions that rotate slow. There is evidence of slow rotation or none among the hypervelocity stars discovered by \cite{2018ApJ...865...15S}. The mechanism to slow down these hypervelocity companions should be studied in depth as well.

\begin{acknowledgments}
  This work is supported by the National Science and Technology Council of Taiwan through grants 113-2112-M-007-031, by the Center for Informatics and Computation in Astronomy (CICA) at National Tsing Hua University through a grant from the Ministry of Education of Taiwan. {\tt FLASH} was in part developed by the DOE NNSA-ASC OASCR Flash Center at the University of Chicago. The simulations and data analysis have been carried out at the {\tt Taiwania-3} supercomputer in the National Center for High-Performance Computing (NCHC) in Taiwan, and on the CICA cluster at National Tsing Hua University. Analysis and visualization of simulation data were completed using the analysis toolkit {\tt yt}.
PR-L acknowledges support from grant PID2021-123528NB-I00, from the
the Spanish Ministry of Science and Innovation
(MICINN). 
JIGH acknowledges financial support from the Spanish Ministry of Science, Innovation and Universities (MICIU) projects PID2020-117493GB-I00 and PID2023-149982NB-I00.

\software{FLASH \citep{2000ApJS..131..273F, 2008PhST..132a4046D}, yt \citep{2011ApJS..192....9T}, Matplotlib \citep{2007CSE.....9...90H}, NumPy \citep{2011CSE....13b..22V}, SciPy \citep{2019zndo...3533894V}}
\end{acknowledgments}

%-------------------------------------------------------------------------------------------------------------
% References
%-------------------------------------------------------------------------------------------------------------

\end{document}